\pdfoutput=1 
\documentclass{aa} 
\usepackage{graphicx,lscape}
\usepackage{txfonts}
\usepackage{color}

\begin{document}
\title{Kinematic analysis and membership status of
TWA22\,AB\thanks{Based on observations performed at the European
Southern Observatory, Chile (76.C-0543, 077.C-0112, 078.C-0158,
079.C-0229).}$^{,}$\thanks{Table 4 is only available in the electronic 
edition of the journal.}} \author{R. Teixeira \inst{1,2}
  \and C. Ducourant \inst{2}
  \and G. Chauvin \inst{3}
  \and A. Krone-Martins \inst{1,2}
  \and M. Bonnefoy \inst{3}
  \and I. Song \inst{4}}

\offprints{teixeira@astro.iag.usp.br}

\institute{
Instituto de Astronomia, Geof\'isica e Ci\^encias Atmosf\'ericas,
       Universidade de S\~ao Paulo,
       Rua do Mat\~ao, 1226 - Cidade Universit\'aria,
       05508-900 S\~ao Paulo - SP,
       Brazil.
          \and
Observatoire Aquitaine des Sciences de l'Univers, CNRS-UMR 5804, BP 89, 33270 Floirac, France.
         \and
	  Laboratoire d'Astrophysique, Observatoire de Grenoble,
          414, Rue de la piscine, 38400 Saint-Martin d'H\`eres, France.
         \and
	  Department of Physics \& Astronomy, the University of
	  Georgia, Athens, GA 30605 USA.}

\date{Received  / Accepted }
\titlerunning {TWA22\,AB kinematics}
\abstract 
{TWA22 was initially regarded as a member of the TW Hydrae
association (TWA). In addition to being one of the youngest ($\approx$
$8$~Myr) and nearest ($\approx$ $20$~pc) stars to Earth, TWA22 has
proven to be very interesting after being resolved as a tight, very
low-mass binary. This binary can serve as a very useful dynamical
calibrator for pre-main sequence evolutionary models.  However,
its membership in the TWA has been recently
questioned despite due to the lack of accurate kinematic measurements.}
{Based on proper motion, radial velocity, and trigonometric
parallax measurements, we aim here to re-analyze the membership of
TWA22 to young, nearby associations.}
{Using the ESO NTT/SUSI2 telescope, we observed TWA22\,AB during 5
different observing runs over 1.2 years to measure its
trigonometric parallax and proper motion. This is a part of a
larger project measuring trigonometric parallaxes and proper
motions of most known TWA members at a sub-milliarcsec level.  HARPS at
the ESO 3.6m telescope was also used to measure the system's radial
velocity over 2 years.}  
{We report an absolute trigonometric parallax of TWA22\,AB,
$\pi=57.0\pm0.7$ mas, corresponding to a distance $17.5\pm0.2$ pc
from Earth. Measured proper motions of TWA 22AB are
$\mu_{\alpha}cos(\delta)=-175.8\pm0.8$ mas/yr and
$\mu_{\delta}=-21.3\pm0.8$ mas/yr. Finally, from HARPS
measurements, we obtain a radial velocity
$V_{rad}=14.8\pm2.1$~km/s.}
{A kinematic analysis of TWA22\,AB space motion and position
implies that a membership of TWA22\,AB to known young, nearby
associations can be excluded except for the $\beta$ Pictoris and
TW Hydrae associations. Membership probabilities based on the
system's Galactic space motion and/or the trace-back technique
support a higher chance of being a member to the $\beta$ Pictoris
association.  Membership of TWA22 in the TWA
cannot be fully excluded because of large uncertainties in parallax
measurements and radial velocities and to the uncertain internal velocity
dispersion of its members.}

\maketitle

\section{Introduction}

Over the past decade, various coeval moving groups of stars have been
discovered in the solar neighborhood, including three well-known
associations, TW Hydrae (hereafter TWA, \cite{kast97}), $\beta$
Pictoris (hereafter $\beta$ Pic, \cite {barr99}), and
Tucana/Horologium (hereafter Tuc-Hor, \cite{zuck00}, \cite{torr00}). 
These groups occupy an important astrophysical
niche thanks to their proximity ($\le100$ pc) and youth ($\le 100$
Myr). They offer the best targets for various studies, such as imaging
searches for young brown dwarf and planetary mass companions,
proto-planetary or debris disk-related programs, etc. Any 
young (sub)stellar binary members in tight orbits in these moving
groups can serve as valuable calibrators for evolutionary models.

\cite{song03} identified the TWA22 system as a new member of TWA
based essentially on the presence of a strong Li $\lambda6708$
absorption feature and its close proximity to TWA in the sky. A further
proper motion analysis led \cite{scho05} to conclude that TWA22
could indeed be the nearest TWA member to Earth. However, more
recently, \cite{mama05} has performed a convergent point analysis
using several TWA members. He found that TWA22 has a low
probability of membership in the TWA. This conclusion was refuted by
\cite{song06} arguing a lack of reliable distance determinations
for most TWA members to firmly reject TWA22's membership.
\cite{song06} argued instead that the strong lithium line
seen at TWA22, rarely seen among other stars with similar
spectral types, implies a probable membership to TWA or
$\beta$ Pic.

During a VLT/NACO deep-imaging survey for close companions to
stars in young associations, TWA22 was resolved as a tight
($\approx$ $100$~mas) binary with a projected physical separation of
$1.76\pm0.10$~AU (see Bonnefoy et al. 2009). 

Regardless of TWA22's membership in TWA or $\beta$ Pic, TWA22\,AB
is a precious dynamical mass calibrator because of its young age
($\approx$ $10$~Myr) and the low masses of its components.  A well-known age
and a distance are crucial to the validity of calibrating
evolutionary model calculations. Therefore, there is need for
accurate observational data, such as the trigonometric parallax,
proper motions, and radial velocity.

As a part of a larger project for the trigonometric parallax
determination of all TWA members, we performed astrometric and
photometric observations of TWA22\,AB at ESO NTT/SUSI2 (La Silla -
Chile).  Radial velocity measurements were also obtained to complete
the kinematic analysis of TWA22\,AB.  Using these new data, we
performed a kinematic analysis to test the membership of TWA22\,AB in
TWA, $\beta$ Pic, and Tuc-Hor. Our analysis uses only moving- group members
with Hipparcos measured distances. In addition to Hipparcos
moving-group members, we used 2M1207A whose accurate
trigonometric distance was obtained from gound-based measurement
(Ducourant et al. 2008). The possibility of TWA22\,AB being a member of other
nearby groups was investigated and discarded by more evident
incompatibilities.

We present observation and data reduction in Sect. 2, and in
Sect. 3, we discuss the membership of TWA22 in TWA, $\beta$ Pic, and
Tuc-Hor.  The conclusion is provided in Sect. 4.

\section{Observations and data treatment}

\subsection{Trigonometric parallax and proper motion}

We used ESO NTT-SUSI2 which provides a good compromise between a large
field of view $(5.5\arcmin \times 5.5\arcmin)$ for a sufficient
sampling of background stars and a small pixel size ($80.5$ mas) necessary for
sub-milli arcsecond astrometry. Data were acquired over five
observational epochs, and all observations were obtained around the
meridian transit (hour angle $\le$ 0.5h). In our imaging, we used an I-band filter to minimize
the differential color refraction effects (DCR).  Residual DCR effects
were removed from single observations following the method described
in \cite {duco07}. Multiple images were obtained at each epoch to
reduce astrometric errors. The image of highest quality is selected as a master frame that will be used for cross-identification, alignement and scaling. The alignment of CCD axes and the plate
scale were estimated using 2MASS catalog sources (\cite {cutr03}).

A stellar point-spread function for each frame was fitted  using
the {\tt DAOPHOT II} package (\cite{stet87}).  Then, we created
catalogs of measured positions $(x,y)$, internal magnitudes, and
associated errors for all stars on each frame. Observational data were
processed through a global treatment, as described in Ducourant et al. (2007, 2008), and a solution was derived for TWA22\,AB relative to
background stars (14.5 $\le$ I $\le$ 18.5 mag). In our astrometry data reduction
and analysis, we ignore any influence of binarity. But, we assessed
the effect of binarity in Sect. 2.2. 
Then, a statistical
conversion from relative to absolute parallax and proper motions,
based on the Besan\c con Galaxy model (\cite{robi03}, 2004),
was derived ($\Delta\pi=0.35\pm0.01$ mas,
$\Delta\mu_{\alpha}cos(\delta)=-5.43\pm0.02$ mas/yr,
$\Delta\mu_{\delta}=+2.98\pm0.02$ mas/yr). Both the final estimated TWA22\,AB ($\alpha = 10^{h} 17^{m} 26.79^{s}$, $\delta = -53\degr 54\arcmin
26.5\arcsec$, epoch = 2006.763 yr)
proper motion and trigonometric parallax are given in Table
\ref{astro}. In a separate table (Table \ref{phot}), apparent and
absolute Bessel (V, R, I) and 2MASS (J, H, K)
magnitudes (\cite{cutr03}) are listed.

\begin{figure}[ht]
  \includegraphics*[angle=0,width=8cm]{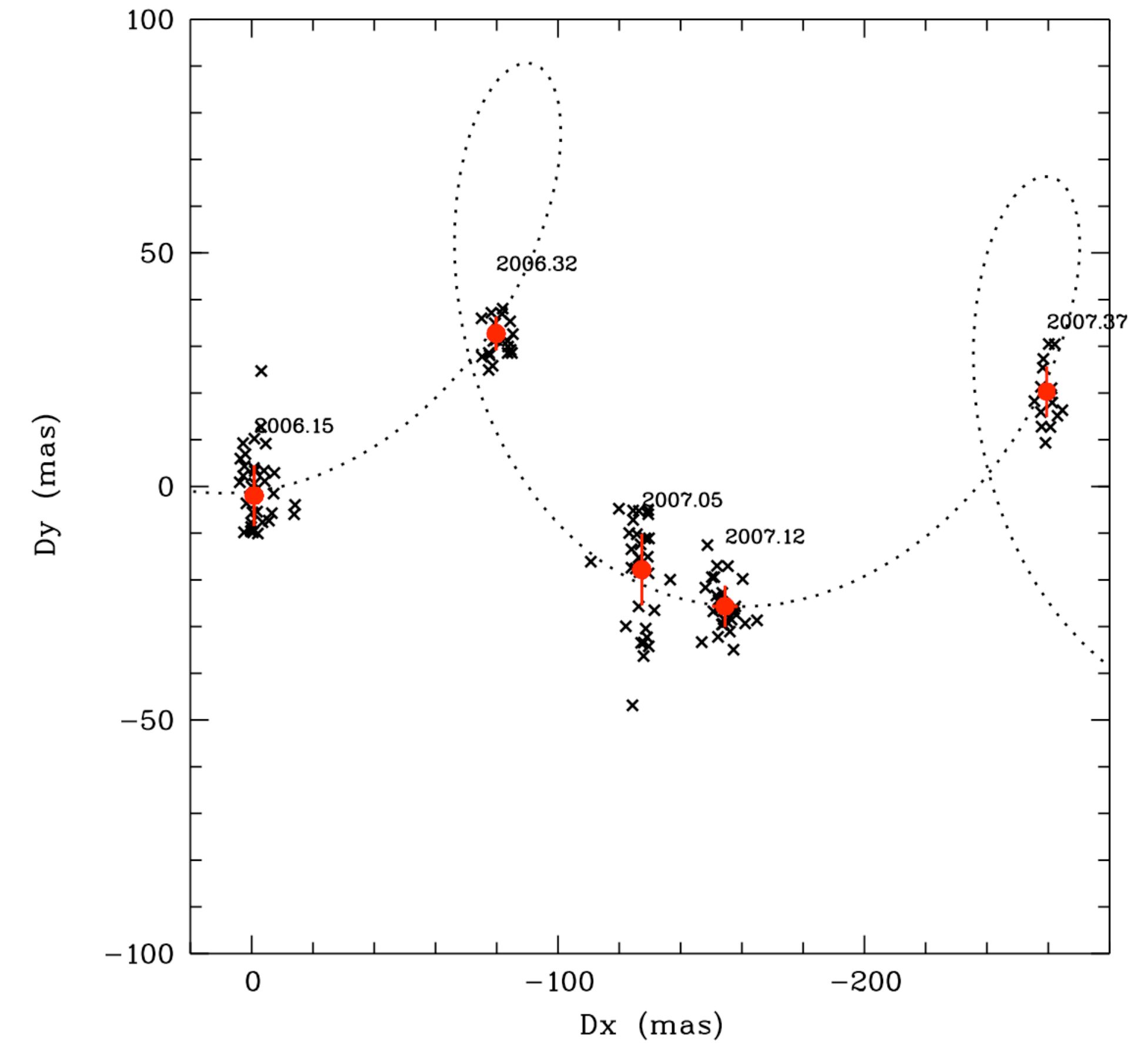}
  \caption{\label{pi}Apparent astrometric displacement of TWA22\,AB
  together with the best parallax and proper motion fit of the data.
  Blue dots correspond to weighted mean positions at each
  observational epoch.}

\end{figure}

Figure \ref{pi} presents the apparent displacement of TWA22\,AB
relative to the background stars due to the parallax and proper
motion.

\begin{table}[ht]
\caption{\label{astro}Absolute astrometric parameters for TWA22\,AB
 and radial velocity derived in this
work.} \centerline{ 
  \begin{tabular}{ccccc}
    \hline\hline
      $\pi$ 	& $d$ & $\mu_{\alpha}cos(\delta)$  & $\mu_{\delta}$ &$V_{rad}$  \\
        mas &pc &mas/yr                             & mas/yr                       & km/s  \\
     \hline\noalign{\smallskip}{\smallskip}
       57.0$\pm$0.7	&17.5$\pm $0.2	&-175.8$\pm$0.8  &-21.3$\pm$0.8 &14.8$\pm$2.1\\
\hline
  \end{tabular}
  }
\end{table}

\begin{table}[ht]
\caption{\label{phot}Apparent and absolute magnitudes of TWA22\,AB: 
V, R, I Bessell filter data from this work and  J, H, and K$_{s}$ data from 2MASS.  }
\centerline{
  \begin{tabular}{rrr}
    \hline\hline
 &\multicolumn{1}{c}{m} & \multicolumn{1}{c}{M}\\
 & \multicolumn{2}{c}{(mag)} \\
     \hline\noalign{\smallskip}{\smallskip}
V & 13.99  $\pm$ 0.02 &  12.77 $\pm$ 0.19\\ 
R & 12.50  $\pm$ 0.14 &  11.28 $\pm$ 0.19\\
I & 11.13  $\pm$ 0.23 &   9.91 $\pm$ 0.18\\
J &  8.55  $\pm$ 0.01 &   7.33 $\pm$ 0.07\\
H &  8.09  $\pm$ 0.04 &   6.87 $\pm$ 0.10\\
K &  7.69  $\pm$ 0.02 &   6.47 $\pm$ 0.07\\
\hline
  \end{tabular}
  }
\end{table}

\subsection{Impact of binarity onto the parallax determination.}

With ESO NTT/SUSI2, it is not possible to resolve two components of
TWA22\,AB. Therefore, with our astrometric observations we are measuring
the photocenter of the system. This photocenter may not coincide
with the center of mass because these two different positions depend
on the ratio of mass and luminosity of the components. A photocenter
of a binary should be affected by an elliptic, periodic movement with
the same period as the binary orbit.  The amplitude of this variation
depends on the orbital parameters, mass, and luminosity ratios.

Following \cite{VdK67}, our condition equations for TWA22\,AB were
modified to include the photocentric movement. The condition
equations are written for each star on each of the N frames considered
(including the master frame).  These equations relate the measured
coordinates to the stellar astrometric parameters:

\begin{equation}
X_{0} +	\Delta X_{0} + \mu_{X} (t-t_{0}) + \pi F_{X}(t) + \alpha Q_{\alpha}= a_{1} x(t) + a_{2} y(t) + a_{3} 
\end{equation}
\begin{equation}
Y_{0} +	\Delta Y_{0} + \mu_{Y} (t-t_{0}) + \pi F_{Y}(t) + \alpha Q_{\delta} = b_{1} x(t) + b_{2} y(t) + b_{3} 
\end{equation}

\begin{table*}[t]\centering

\caption{\label{UVW}Heliocentric coordinates (X, Y, Z) and
Galactic velocities (U, V, W) of TWA22\,AB together with mean
values for TWA, $\beta$ Pic and Tuc-Hor.}

\begin{tabular}{lcccccccc}
\hline\noalign{\smallskip}{\smallskip}
    Name & X & Y & Z &  U         & V           & W                 & Age & d\\
         &   & pc &  &            &  km/s       &                   & Myr  & pc\\
\hline
\hline
TWA22       &   3.5 $\pm$0.2   & -17.2$\pm$0.2   &     0.7$\pm$ 0.2 &  -8.0$\pm$0.4 & -17.1$\pm$2.1 & -9.0$\pm$0.2     &  $\le$10(1)  & 17.5$\pm$0.2\\ 
\hline
TWA           & 14.4$\pm$8.7   & -46.9$\pm$5.3 &   22.5$\pm$2.5  &  -10.0$\pm$2.0 & -17.6$\pm$1.4 & -4.8$\pm$1.1      &8(2)	& 55$\pm$7 \\
$\beta$ Pic & 9.2$\pm$28.9   &   -7.5$\pm$13.0 & -13.2$\pm$6.5 & -10.9$\pm$1.9 & -16.1$\pm$1.0 &-9.0$\pm$1.2        &12 (3)  &34$\pm$14 \\
Tuc-Hor      &14.1$\pm$17.7 &   -19.4$\pm$8.0 & -34.9$\pm$3.5 & -9.5$\pm$1.7 & -20.6$\pm$1.7 &-0.6$\pm$2.6      &27 (4)  &46$\pm$5 \\
\hline
\end{tabular}
\end{table*}

\noindent where ($X_{0}, Y_{0}$) are the known standard coordinate of
the star at the central epoch $t_0$ and ($x(t), y(t)$)
are the measured coordinates on the frame (epoch $t$) that need to be
transformed into the master frame system.  $\Delta X_{0}$, $\Delta
Y_{0}$, $\mu_{X}$, $\mu_{Y}$, $\pi$, and $\alpha$ are the unknown
stellar astrometric parameters. Both $\Delta X_{0}$ and $\Delta Y_{0}$
yield corrections of the standard coordinates of the star on the
master frame, $\mu_{X}$ and $\mu_{Y}$ are projected proper motions in
right ascension and declination, $\pi$ is the parallax, and $\alpha$
is a semi--major axis of the photocentric trajectory relative to a
barycenter.  Coefficients ($a_{i}$, $b_i$) are unknown frame
parameters that describe the transformation to the master frame
system,  $(F_{\alpha},F_{\delta})$ are the parallax factors, and
$(Q_{\alpha},Q_{\delta})$ are known orbital factors (based on
\cite{bonn08} orbital data).
The unknown coefficient $\alpha$ is given by $\alpha=a(R-\beta)$
where $a$ is the semi-major axis of TWA22\,B's orbit around TWA22\,A,
R is a fractional mass, $R=\frac{M_{B}}{M_{A}+M_{B}}$,  $\beta$ is the
fractional distance of the primary to the photocenter
$\beta=\frac{1}{1+10^{-0.4\Delta m}}$ where $\Delta m$ is the
magnitude difference between A and B.

Although it is impossible to include the parameter $\alpha$ formally as
a variable in our equations, we could externally determine its value
to correct $X_{0}, Y_{0}$ for the orbital motion of
photocenter around barycenter. For this, we assumed the
mass--luminosity relations as: $M~\propto~(L)^{-2.5}$ and used the
extrapolated magnitudes in the I-band adapted from \cite{bonn08}: $\Delta
m_{I}=0.46-0.87$ mag to determine $\alpha Q_{\alpha}$ and $\alpha
Q_{\delta}$ and correct $X_{0}, Y_{0}$ from these quantities. The
resulting astrometric parameters for TWA22\,AB are then:
$\pi=56.4\pm0.7$ mas, $\mu_{\alpha}cos(\delta)=-178.2\pm0.7$
mas/yr, and $\mu_{\delta}=-9.4\pm0.8$ mas/yr.

We note that the parallax value obtained from this study is within $1
\sigma_{\pi}$ of the one given in Table \ref{astro} where it did
not consider the effect of photocenter's periodic movement. On the
other hand, we observe a large variation in the proper motion in
declination. This suggests that our astrometric data may not cover a
long enough time interval to properly account for the 5.144\,yr, orbital
periodic signal from \cite{bonn08}. The influence of this
signal over 1.2 yr (duration of the observational program of TWA22\,AB)
is mainly equivalent to an offset of the proper motion in declination.
We therefore conservatively stick to the values given in Table
\ref{astro} as the best fit to our data.

\subsection{Radial velocity}

TWA22 AB was observed with HARPS (\cite{mayor03}) over 2 years.  HARPS
is a high-resolution (R=115,000) fiber-fed cross-dispersed echelle
spectrograph functioning on the ESO/3.6m telescope. For TWA22 AB
($V=13.99$), we used 15 min exposures to obtain 60 spectra with SNR
ranging from 6 to 10. The standard HARPS reduction pipeline was used
to derive radial velocities from the cross-correlation of spectra with
a mask for an M2 star. The instrument is generally stable over one
night (nightly instrumental drifts $\le 1$~m.s$^{-1}$ ), and we
performed a precise nightly wavelength calibration from ThAr spectra
(\cite{lovi07}).  Out of these measurements, we estimated a
heliocentric radial velocity $V_{r}$ = 14.8 $\pm 2.1$~km.s$^{-1}$ for
the system.  This radial velocity is likely to be affected by
the SB1 status of TWA22\,AB. We took the binary nature into account in
our uncertainty value by simulating expected amplitudes from the
binary orbital motion based on the parameters from Bonnefoy et al.
(2009).

\section{Kinematic analysis}

Based on new trigonometric parallax, proper motion, and radial
velocity, we re-examine below the membership of TWA22\,AB to TWA,
$\beta$ Pic, and Tuc--Hor. Membership in other young, nearby
associations was rejected by more evident discrepancies in age,
distances, etc.

\subsection{Space motion}

To test the membership of TWA22\,AB, our first approach is to
statistically compare its galactic space motion with the mean UVW
values for TWA, $\beta$ Pic, and Tuc-Hor members.  Only stars with
known trigonometric parallaxes were considered here because accurate
distances are crucial in the UVW calculation.  For TWA, our sample set
includes TWA\,01, TWA\,04, TWA\,09, and TWA\,11 with trigonometric
parallaxes from the Hipparcos catalog (\cite{esa97})and 2M1207A from
\cite{duco08}. Although TWA19 has a Hipparcos parallax, we did not consider it here because \cite{mama05} classified it as non TWA member. SSSPMJ1102-3431 TWA member (\cite{teix08}) has also been excluded from our study because no radial velocity was available. Their radial velocities were extracted from Table~1 of
\cite{mama05} (\cite{torr03} data for TWA\,01 and TWA9A; \cite{torr95}
for TWA4; \cite{reid03} for TWA11; and \cite{mohanty03} for 2M1207A). For
$\beta$ Pic, we used the list of suggested members by
\cite{torr06}.  Astrometric measurements and radial velocities
were respectively obtained from the Hipparcos catalog
(\cite{esa97}) and the Table~6 of \cite{torr06}. Finally, for
Tuc-Hor, astrometric measurements and radial velocities were
respectively obtained from the Hipparcos catalog
(\cite{esa97}) and \cite{khar07} for suggested members by
\cite{zuck04} and \cite{torr08}.  The selection of stars and
the original data used in our analysis are presented in
Table~\ref{data} (only available in the electronic edition). Calculated mean space
motions of TWA, $\beta$ Pic, and Tuc-Hor are reported in
Table~\ref{UVW}, together with their spatial heliocentric
coordinates and velocities for TWA22\,AB. In this Table  positive X(U) points to the
Galactic center, Y(V) is positive in the direction of Galactic
rotation and Z(W) is positive toward the north Galactic pole.  Ages
are from (1) \cite{song03}, (2) \cite{dela06}, (3) \cite{orte04}, (4) \cite{maka07}.

The mean values
derived here (Table~\ref{UVW}) for TWA, $\beta$ Pic, and Tuc-Hor are in good agreement with
published values : (-11, -18, -5), (-11, -16, -9), and (-11,
-21, 0) from Table 7 of \cite{zuck04} and ($-10.5\pm0.9, -18.0\pm1.5,
-4.9\pm0.9$), ($-10.1\pm2.1, -15.9\pm0.8, -9.2\pm1.0$), and
($-9.9\pm1.5,-20.9\pm0.0, -1.4\pm0.9$) from Table 1 of
\cite{torr08}.

Since Hipparcos proper motions were derived from observations covering
only a small time interval, we also calculated mean spatial motions of
these associations using Tycho-2 proper motions (\cite{hoeg00}) and we
obtained nearly the same results: (-10.1, -17.3, -5.0),
(-10.8,-16.0,-9.1), and (-9.6, -20.8, -0.5), respectively, for TWA,
$\beta$ Pic, and Tuc-Hor.

To test the membership of TWA22\,AB to TWA, $\beta$ Pic, and Tuc-Hor,
based on its space motion, we applied a $\chi^{2}$ test with 3 degrees
of freedom using their space motion measurements. We find the
probabilities that TWA22\,AB space motion be compatible with the mean
space motion of $\beta$ Pic, TWA and Tuc-Hor are 50\%, 1\% and 0.5\%
respectively. An alternative approach is to perform a k-NN analysis in
the UVW space (implemented by \cite{vena02}, \cite{R08}) where we computed the
distance of TWA22 to all members of these associations. Among the k
nearest neighbors to TWA22, the fraction of members for a given
association gives the membership probability for that group. This k-NN
analysis corroborates that TWA22 is more likely a member of $\beta$
Pic than TWA and Tuc-Hor. Both calculations tend to reject TWA and
Tuc-Hor as a host association for TWA22\,AB.

\begin{figure}[t]
  \includegraphics*[angle=0,width=8cm]{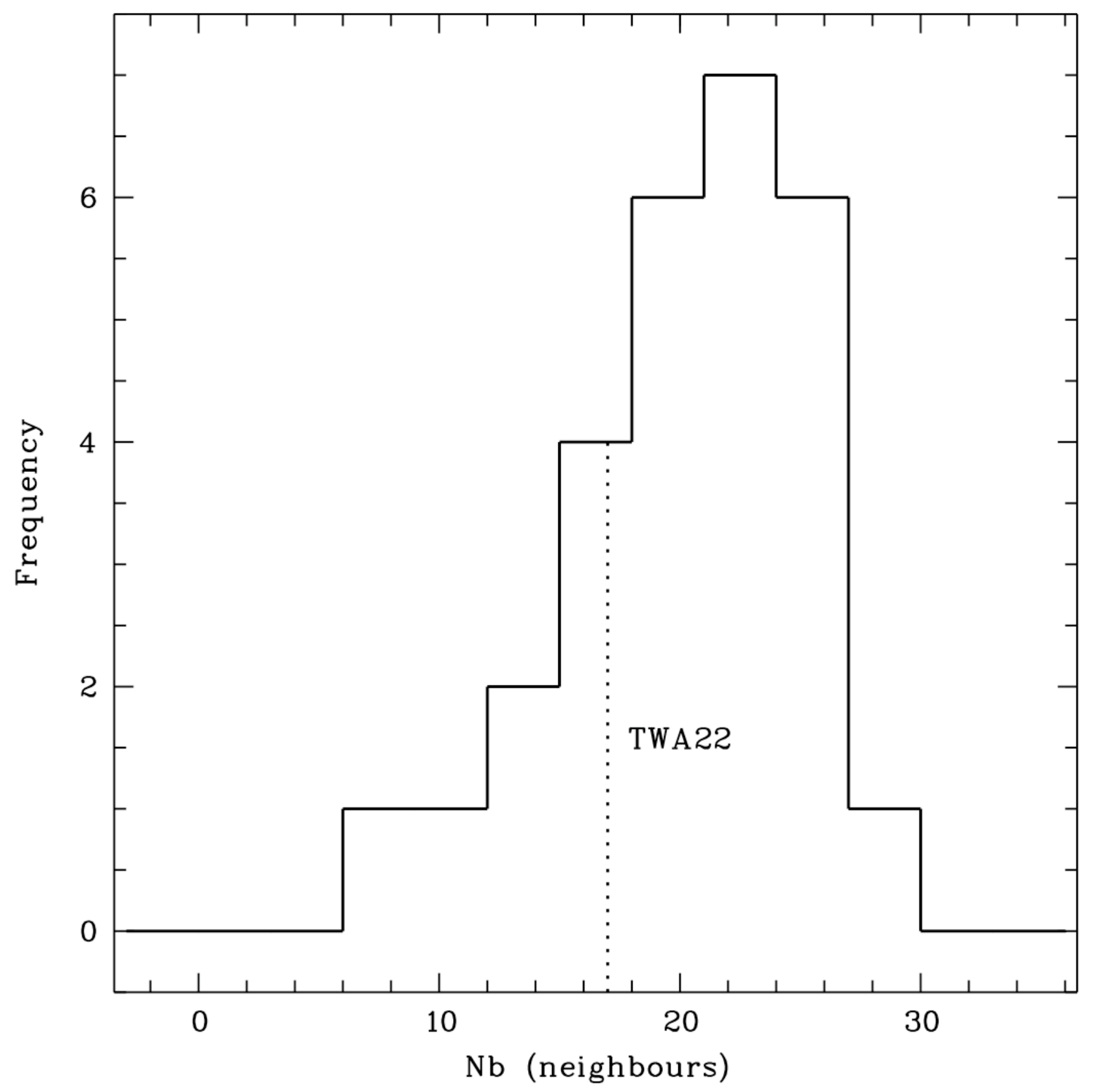}

  \caption{\label{voisinage} Number of $\beta$ Pic members in a sphere
  of 2$\sigma$ radius in the UVW space around each member. Sigma is
  the velocity dispersion for this association.}
\end{figure}

Our final approach is to count the number of $\beta$ Pic, TWA,
and Tuc-Hor members (or neighbors) within a sphere of fixed radius
in the UVW space, centered at TWA22's value and at each individual member of these
associations. Radii are selected proportional to the velocity
dispersions of the associations. We found that UVW spatial densities of TWA
and Tuc-Hor members around TWA22 are significantly lower than the
averaged one found in both associations. In contrary, the UVW
spatial density of $\beta$ Pic members around TWA22 (within
2$\sigma$) is similar to the average density in $\beta$ Pic (see Fig.
\ref{voisinage}). 

Membership probabilities and association member density around TWA22
support a possible membership in the $\beta$ Pic association. The
membership of TWA22 in TWA cannot be firmly evaluated because of
the paucity of high-quality kinematic data for most of its members.

\subsection{Trace-back}

\begin{figure}[t]
  \includegraphics*[angle=0,width=8cm]{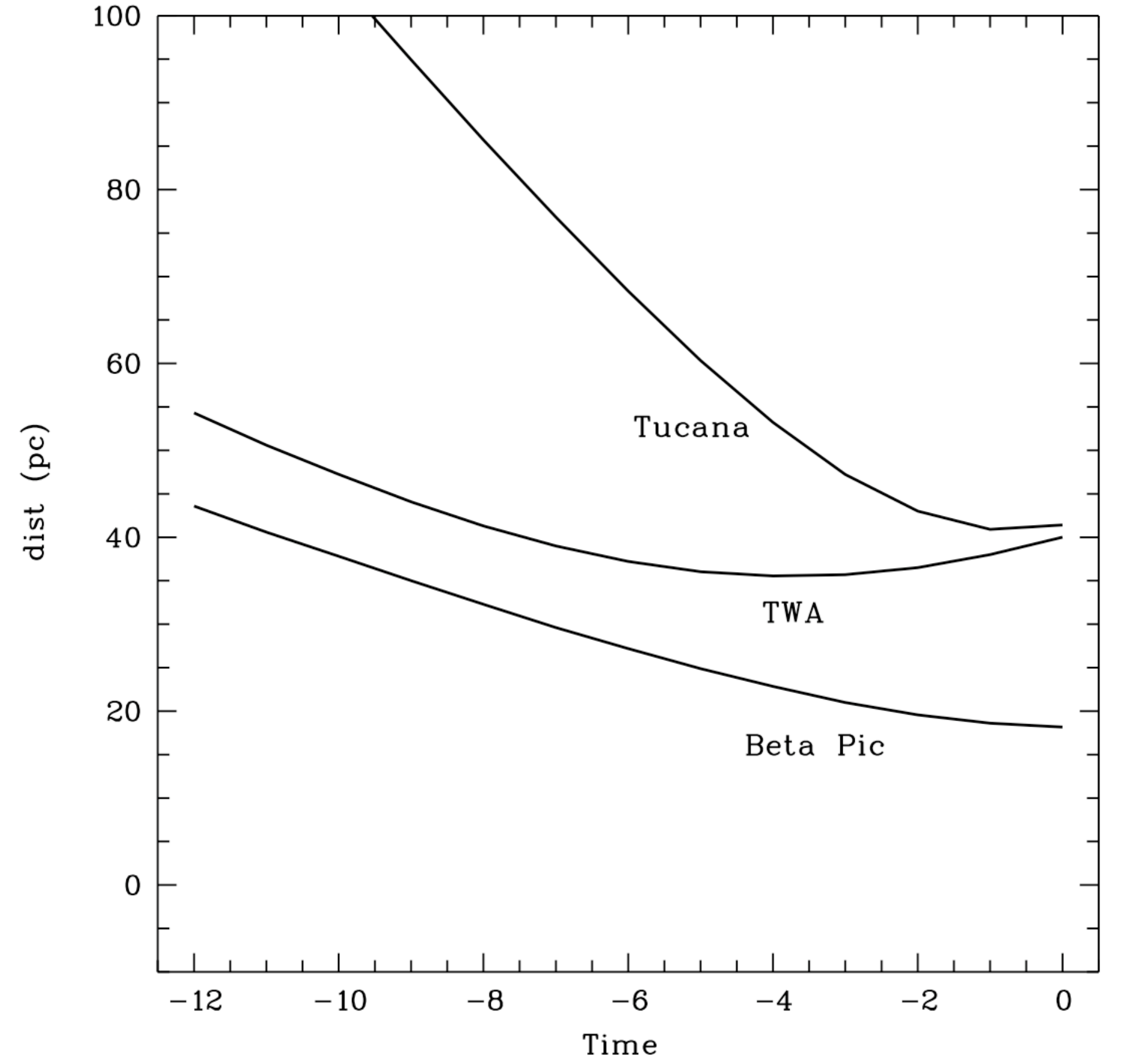}
  \caption{\label{dist}Distances between TWA22\,AB and three moving
  groups (TWA, $\beta$ Pictoris, and Tucana-Horologium moving
  groups) for the past 12 Myrs.} \end{figure}

Another way to test the membership of TWA22\,AB to TWA, $\beta$ Pic,
and Tuc-Hor is to use the trace-back technique to compare the Galactic
space position of TWA22\,AB and TWA, $\beta$ Pic, and Tuc-Hor backward
in time. To avoid using any uncertain values for the solar peculiar
motion (\cite{miha81}; \cite{dehn98}; Makarov 2007) by transforming
heliocentric velocities into LSR velocities, we decided to work in a
reference system centered, along time, on TWA22\,AB instead. We
present, in Figure~\ref{dist}, the distance between TWA22\,AB and the
center of each association in time.  We note that TWA22 is always
closer to $\beta$ Pic than to TWA and Tuc-Hor.  Since the mean motions
of the associations derived here are in good agreement with values
from the literature, the few TWA members considered here
may not have seriously affected the result.

\section{Conclusions}

Motivated by the importance of the young, very low-mass astrometric
binary TWA22\,AB as an important calibration point for stellar
theoretical calculations, we measured its precise trigonometric
parallax ($57.0\pm0.7$ mas), proper motions ($-175.8\pm0.8$
mas/yr,$-21.3\pm0.8$ mas/yr), and radial velocity
($V_{rad}=14.8\pm2.1$~km/s). These parameters are fundamentals for
determin the physical properties of the tight binary system
(Bonnefoy et al. 2009). 

Our high-quality astrometric measurements along with HARPS radial
velocity measurement allow us to discuss the membership of TWA22\,AB
to nearby associations.  Our
kinematical study shows that membership by TWA22\,AB in known young, nearby
associations can be excluded except for the $\beta$ Pictoris and TW
Hydrae associations. Membership probabilities based on the system
space motion or the use of trace-back technique also support possible
membership of TWA22\,AB in the $\beta$ Pictoris association.
Membership in the TWA cannot be fully excluded because of
the current lack of precise parallax measurements for most of its
members. Our results are, to some extent, inconclusive about the membership
of TWA22\,AB in TWA or $\beta$ Pic, but they are consistent with that from an
age analysis. The location of TWA22\,AB on a color-magnitude diagram
supports its age being about 10\,Myr but cannot be determined 
precisely enough to distinguish from 8 (TWA age) and 12 Myr ($\beta$ Pic
moving group age). Precisely known trigonometric distances of many
more TWA members, an aim of our larger astrometric program of
observing all known TWA members, should improve the situation soon.

\begin{acknowledgements} 
We thank Michel Rapaport for helpful comments.  We also acknowledge
partial financial support from the {\sl Programmes Nationaux de
Plan\'etologie et de Physique Stellaire} (PNP \& PNPS) (in France),
the Brazilian organization FAPESP and CAPES, and the French organization COFECUB.
\end{acknowledgements}

\begin{landscape}
\begin{table}[ht] \centering

\caption{\label{data} Original data for $\beta$ Pictoris, TWA and Tucana/Horologium members used in our kinematical analysis from \cite{esa97}, \cite{mama05}, \cite{torr06} and \cite{duco08} together with the derived heliocentric spatial coordinates and velocities derived in this work. } 
  \begin{tabular}{lrrrrrrrrrrrrr}

\hline
    \multicolumn{1}{c}{ident} & \multicolumn{1}{c}{alpha} & \multicolumn{1}{c}{delta} & \multicolumn{1}{c}{V} &  \multicolumn{1}{c}{$\pi$} &  \multicolumn{1}{c}{$\mu_{\alpha}cos(\delta)$}  &  \multicolumn{1}{c}{$\mu_{\delta}$} &  \multicolumn{1}{c}{Vr} &  \multicolumn{1}{c}{X} &  \multicolumn{1}{c}{Y} &  \multicolumn{1}{c}{Z} &  \multicolumn{1}{c}{U} &  \multicolumn{1}{c}{V} &  \multicolumn{1}{c}{W}\\
    & \multicolumn{1}{c}{(hms)} & \multicolumn{1}{c}{($\degr \arcmin \arcsec$)} & \multicolumn{1}{c}{mag} &  \multicolumn{1}{c}{(mas)} &  \multicolumn{2}{c}{(mas/yr)}  &  \multicolumn{1}{c}{(Km/s)} &  \multicolumn{3}{c}{(pc)}  &  \multicolumn{3}{c}{(Km/s)} \\
      \hline\\
      
HIP10679    & 02 17 24.68 & +28 44 31.0 &  7.75 &  29.40$\pm$5.39 &    98.15$\pm$6.96 &   -67.41$\pm$5.77 &   5.0$\pm$0.4 &  -24.0 &   16.9 &  -17.2 &  -12.6$\pm$   1.9 &  -14.0$\pm$   3.2 &   -6.0$\pm$   1.1 \\ 
HD 14082    & 02 17 25.23 & +28 44 42.8 &  6.99 &  25.37$\pm$2.84 &    94.34$\pm$3.00 &   -72.17$\pm$2.83 &   4.6$\pm$0.3 &  -27.8 &   19.6 &  -20.0 &  -13.2$\pm$   1.2 &  -16.9$\pm$   2.2 &   -7.3$\pm$   0.7 \\ 
AG Tri      & 02 27 29.20 & +30 58 25.2 & 10.12 &  23.66$\pm$2.04 &    79.50$\pm$2.26 &   -70.09$\pm$1.54 &   7.0$\pm$1.1 &  -31.1 &   20.9 &  -19.5 &  -14.0$\pm$   1.2 &  -15.0$\pm$   1.7 &   -8.8$\pm$   0.8 \\ 
BD+05 378   & 02 41 25.84 & +05 59 18.9 & 10.28 &  24.67$\pm$2.41 &    82.32$\pm$4.30 &   -55.14$\pm$2.72 &  10.0$\pm$0.0 &  -26.5 &	6.7 &  -29.9 &  -12.1$\pm$   0.8 &  -16.5$\pm$   1.9 &   -6.6$\pm$   0.4 \\ 
HD 29391    & 04 37 36.11 & -02 28 24.2 &  5.22 &  33.60$\pm$0.91 &    43.32$\pm$0.81 &   -64.23$\pm$0.61 &  21.0$\pm$0.0 &  -24.3 &   -8.2 &  -15.2 &  -14.0$\pm$   0.1 &  -16.2$\pm$   0.3 &  -10.1$\pm$   0.1 \\ 
GJ 3305*    & 04 37 36.11 & -02 28 24.2 &  5.22 &  33.60$\pm$0.91 &    43.32$\pm$0.81 &   -64.23$\pm$0.61 &  20.1$\pm$0.0 &  -24.3 &   -8.2 &  -15.2 &  -13.2$\pm$   0.1 &  -16.0$\pm$   0.3 &   -9.6$\pm$   0.1 \\ 
V1005 Ori   & 04 59 34.81 & +01 47 01.5 & 10.05 &  37.50$\pm$2.56 &    37.15$\pm$2.18 &   -93.94$\pm$1.48 &  18.7$\pm$0.0 &  -23.3 &   -7.4 &  -10.7 &  -11.7$\pm$   0.3 &  -16.9$\pm$   0.8 &   -9.4$\pm$   0.3 \\ 
CD-57 1054  & 05 00 47.09 & -57 15 26.1 & 10.02 &  38.08$\pm$1.07 &    35.64$\pm$1.09 &    72.80$\pm$0.98 &  19.4$\pm$0.3 &   -1.5 &  -20.8 &  -15.9 &  -10.8$\pm$   0.3 &  -16.7$\pm$   0.3 &   -9.2$\pm$   0.2 \\ 
HIP23418    & 05 01 58.79 & +09 58 59.9 & 11.45 &  31.20$\pm$8.56 &    17.18$\pm$8.75 &   -81.96$\pm$5.42 &  17.3$\pm$0.0 &  -29.8 &   -5.5 &  -10.4 &  -12.4$\pm$   1.1 &  -14.4$\pm$   3.3 &  -10.0$\pm$   1.7 \\ 
HD 35850    & 05 27 04.75 & -11 54 03.0 &  6.30 &  37.26$\pm$0.84 &    17.19$\pm$0.69 &   -49.30$\pm$0.64 &  22.8$\pm$0.0 &  -20.2 &  -13.8 &  -10.9 &  -13.2$\pm$   0.1 &  -17.1$\pm$   0.1 &   -9.9$\pm$   0.1 \\ 
Beta Pic    & 05 47 17.08 & -51 04 00.2 &  3.85 &  51.87$\pm$0.51 &	4.65$\pm$0.53 &    81.96$\pm$0.61 &  20.2$\pm$0.4 &   -3.3 &  -16.3 &	-9.8 &  -10.9$\pm$   0.1 &  -16.2$\pm$   0.3 &   -9.2$\pm$   0.2 \\ 
AO Men      & 06 18 28.22 & -72 02 42.1 &  9.95 &  25.99$\pm$1.02 &    -8.14$\pm$0.98 &    71.42$\pm$1.15 &  16.3$\pm$0.0 &    7.4 &  -33.1 &  -18.2 &   -9.7$\pm$   0.5 &  -16.3$\pm$   0.1 &   -8.8$\pm$   0.2 \\ 
HD 139084B  & 15 38 57.60 & -57 42 26.4 &  8.14 &  25.15$\pm$1.09 &   -52.87$\pm$1.11 &  -105.99$\pm$1.03 &   0.1$\pm$2.0 &   32.1 &  -23.5 &	-1.3 &  -11.9$\pm$   1.7 &  -15.9$\pm$   1.4 &  -10.1$\pm$   0.5 \\ 
V343 Nor*   & 15 38 57.60 & -57 42 26.4 &  8.14 &  25.15$\pm$1.09 &   -52.87$\pm$1.11 &  -105.99$\pm$1.03 &   4.2$\pm$1.4 &   32.1 &  -23.5 &	-1.3 &   -8.6$\pm$   1.2 &  -18.3$\pm$   1.1 &  -10.3$\pm$   0.5 \\ 
V824 Ara    & 17 17 25.54 & -66 57 02.5 &  6.87 &  31.83$\pm$0.74 &   -21.84$\pm$0.58 &  -136.47$\pm$0.64 &   5.9$\pm$0.0 &   24.7 &  -17.3 &	-8.8 &   -8.1$\pm$   0.3 &  -17.5$\pm$   0.3 &   -9.3$\pm$   0.2 \\ 
HD 155555C* & 17 17 25.54 & -66 57 02.5 &  6.87 &  31.83$\pm$0.74 &   -21.84$\pm$0.58 &  -136.47$\pm$0.64 &   2.7$\pm$1.8 &   24.7 &  -17.3 &	-8.8 &  -10.6$\pm$   1.4 &  -15.8$\pm$   1.1 &   -8.4$\pm$   0.5 \\ 
HD 164249   & 18 03 03.41 & -51 38 55.7 &  7.01 &  21.34$\pm$0.86 &	3.46$\pm$0.86 &   -86.46$\pm$0.56 &   0.5$\pm$0.4 &   43.2 &  -14.3 &  -11.3 &   -7.0$\pm$   0.5 &  -15.5$\pm$   0.6 &   -9.0$\pm$   0.4 \\ 
HD 164249B* & 18 03 03.41 & -51 38 55.7 &  7.01 &  21.34$\pm$0.86 &	3.46$\pm$0.86 &   -86.46$\pm$0.56 &  -2.4$\pm$1.3 &   43.2 &  -14.3 &  -11.3 &   -9.6$\pm$   1.2 &  -14.6$\pm$   0.7 &   -8.3$\pm$   0.5 \\ 
HD 168210   & 18 19 52.21 & -29 16 32.4 &  8.72 &  13.25$\pm$1.41 &	2.84$\pm$1.83 &   -47.16$\pm$1.17 &  -7.0$\pm$2.6 &   74.8 &	4.4 &	-8.8 &   -7.1$\pm$   2.6 &  -15.0$\pm$   1.6 &   -7.8$\pm$   1.1 \\ 
HD 172555   & 18 45 26.86 & -64 52 15.2 &  4.78 &  34.21$\pm$0.68 &    32.67$\pm$0.51 &  -148.72$\pm$0.45 &   3.8$\pm$0.0 &   23.3 &  -13.1 &  -11.8 &   -9.5$\pm$   0.3 &  -16.4$\pm$   0.3 &  -10.0$\pm$   0.2 \\ 
CD-64 1208* & 18 45 26.86 & -64 52 15.2 &  4.78 &  34.21$\pm$0.68 &    32.67$\pm$0.51 &  -148.72$\pm$0.45 &   1.0$\pm$3.0 &   23.3 &  -13.1 &  -11.8 &  -11.8$\pm$   2.4 &  -15.1$\pm$   1.4 &   -8.9$\pm$   1.2 \\ 
PZ Tel      & 18 53 05.86 & -50 10 49.1 &  8.43 &  20.14$\pm$1.18 &    16.64$\pm$1.32 &   -83.58$\pm$0.87 &  -3.4$\pm$0.7 &   45.1 &  -11.1 &  -17.6 &  -10.6$\pm$   0.8 &  -15.4$\pm$   1.0 &   -7.9$\pm$   0.7 \\ 
Eta Tel     & 19 22 51.18 & -54 25 25.4 &  5.03 &  20.98$\pm$0.68 &    25.57$\pm$0.75 &   -83.03$\pm$0.49 &   0.0$\pm$0.0 &   40.9 &  -12.6 &  -21.1 &   -9.0$\pm$   0.3 &  -15.4$\pm$   0.5 &   -8.2$\pm$   0.3 \\ 
HD 181327   & 19 22 58.92 & -54 32 16.3 &  7.04 &  19.77$\pm$0.81 &    23.84$\pm$0.89 &   -81.77$\pm$0.57 &  -0.7$\pm$0.0 &   43.3 &  -13.4 &  -22.4 &   -9.9$\pm$   0.4 &  -16.0$\pm$   0.7 &   -8.0$\pm$   0.4 \\ 
AT MicN     & 20 41 50.97 & -32 26 03.6 & 10.27 &  97.80$\pm$4.65 &   269.32$\pm$6.55 &  -365.69$\pm$4.65 &  -4.5$\pm$0.0 &    8.1 &	1.6 &	-6.1 &  -10.2$\pm$   0.4 &  -17.0$\pm$   0.8 &  -10.5$\pm$   0.7 \\ 
AT MicS     & 20 41 50.97 & -32 26 03.6 & 10.27 &  97.80$\pm$4.65 &   269.32$\pm$6.55 &  -365.69$\pm$4.65 &  -5.2$\pm$0.0 &    8.1 &	1.6 &	-6.1 &  -10.8$\pm$   0.4 &  -17.1$\pm$   0.8 &  -10.1$\pm$   0.7 \\ 
AU Mic      & 20 45 09.34 & -31 20 24.1 &  8.81 & 100.59$\pm$1.35 &   280.37$\pm$1.58 &  -360.09$\pm$0.98 &  -6.0$\pm$1.7 &    7.8 &	1.7 &	-6.0 &  -11.3$\pm$   1.3 &  -16.7$\pm$   0.4 &   -9.6$\pm$   1.0 \\ 
WW PsA      & 22 44 57.84 & -33 15 00.7 & 11.70 &  42.35$\pm$3.37 &   183.12$\pm$2.50 &  -118.87$\pm$2.21 &   2.2$\pm$0.0 &   10.8 &	2.4 &  -20.9 &  -12.6$\pm$   1.1 &  -17.9$\pm$   1.5 &  -11.1$\pm$   0.7 \\ 
\\
  \hline\\
TWA01       & 11 01 51.95 & -34 42 16.9 & 10.92 &  17.72$\pm$2.21 &   -66.90$\pm$1.78 &   -12.36$\pm$1.42 &  12.7$\pm$0.2 &    7.8 &  -51.4 &	22.0 &  -12.0$\pm$   1.8 &  -18.0$\pm$   0.8 &   -5.1$\pm$   1.3 \\ 
TWA04       & 11 22 05.34 & -24 46 39.5 &  8.89 &  21.43$\pm$2.86 &   -85.45$\pm$1.89 &   -33.37$\pm$2.12 &   9.3$\pm$1.0 &    5.7 &  -38.4 &	26.0 &  -11.8$\pm$   1.8 &  -17.7$\pm$   1.6 &   -6.8$\pm$   1.7 \\ 
TWA09A      & 11 48 24.26 & -37 28 49.0 & 11.13 &  19.87$\pm$2.38 &   -54.10$\pm$2.12 &   -19.97$\pm$1.54 &   9.5$\pm$0.4 &   15.2 &  -43.5 &	20.2 &   -6.6$\pm$   1.2 &  -14.9$\pm$   0.9 &   -3.5$\pm$   1.0 \\ 
TWA11       & 12 36 01.07 & -39 52 10.0 &  5.78 &  14.91$\pm$0.75 &   -55.92$\pm$0.70 &   -24.00$\pm$0.52 &   6.9$\pm$1.0 &   30.6 &  -53.7 &	26.1 &  -10.1$\pm$   0.8 &  -17.0$\pm$   1.0 &   -5.4$\pm$   0.6 \\ 
2M1207      & 12 07 33.46 & -39 32 53.9 & 20.15 &  19.10$\pm$0.40 &   -64.20$\pm$0.40 &   -22.60$\pm$0.40 &  11.2$\pm$2.0 &   19.5 &  -44.2 &	20.1 &   -7.9$\pm$   0.8 &  -18.3$\pm$   1.7 &   -3.5$\pm$   0.8 \\ 
\\ 
  \hline\\
  
HIP490     & 00 05 52.47 & -41 45 10.4 &  7.51 &  24.85$\pm$0.92 &    97.62$\pm$0.59 &   -76.40$\pm$0.73 &   1.6$\pm$1.3 &   10.6 &   -5.5 &  -38.4 &   -9.8$\pm$   0.5 &  -21.5$\pm$   0.8 &   -1.3$\pm$   1.2 \\ 
HIP1113    & 00 13 52.83 & -74 41 17.4 &  8.76 &  22.86$\pm$0.87 &    84.14$\pm$0.81 &   -47.52$\pm$0.78 &   8.8$\pm$0.2 &   19.2 &  -26.1 &  -29.4 &   -9.0$\pm$   0.5 &  -19.9$\pm$   0.6 &   -1.3$\pm$   0.3 \\ 
HIP1481     & 00 18 26.01 & -63 28 38.5 &  7.46 &  24.42$\pm$0.68 &    90.37$\pm$0.60 &   -58.98$\pm$0.67 &   6.6$\pm$0.3 &   15.4 &  -19.0 &  -32.8 &   -9.0$\pm$   0.4 &  -20.0$\pm$   0.5 &   -0.8$\pm$   0.3 \\ 

  \end{tabular}
\end{table}
\end{landscape}

\begin{landscape}
\begin{table}[ht] \centering
 
  \begin{tabular}{lrrrrrrrrrrrrr}

\hline
    \multicolumn{1}{c}{ident} & \multicolumn{1}{c}{alpha} & \multicolumn{1}{c}{delta} & \multicolumn{1}{c}{V} &  \multicolumn{1}{c}{$\pi$} &  \multicolumn{1}{c}{$\mu_{\alpha}cos(\delta)$}  &  \multicolumn{1}{c}{$\mu_{\delta}$} &  \multicolumn{1}{c}{Vr} &  \multicolumn{1}{c}{X} &  \multicolumn{1}{c}{Y} &  \multicolumn{1}{c}{Z} &  \multicolumn{1}{c}{U} &  \multicolumn{1}{c}{V} &  \multicolumn{1}{c}{W}\\
    & \multicolumn{1}{c}{(hms)} & \multicolumn{1}{c}{($\degr \arcmin \arcsec$)} & \multicolumn{1}{c}{mag} &  \multicolumn{1}{c}{(mas)} &  \multicolumn{2}{c}{(mas/yr)}  &  \multicolumn{1}{c}{(Km/s)} &  \multicolumn{3}{c}{(pc)}  &  \multicolumn{3}{c}{(Km/s)} \\
      \hline\\
HIP2484    & 00 31 32.56 & -62 57 29.1 &  4.36 &  23.35$\pm$0.52 &    82.48$\pm$0.42 &   -54.37$\pm$0.47 &  14.0$\pm$3.0 &   15.1 &  -20.2 &  -34.7 &   -5.3$\pm$   1.1 &  -23.1$\pm$   1.5 &   -6.2$\pm$   2.4 \\ 
HIP2487    & 00 31 33.36 & -62 57 55.6 &  4.53 &  18.95$\pm$4.35 &    87.95$\pm$4.06 &   -45.79$\pm$3.96 &   9.8$\pm$0.0 &   18.6 &  -24.8 &  -42.7 &  -11.1$\pm$   3.5 &  -24.1$\pm$   4.6 &   -3.0$\pm$   1.3 \\ 
HIP2578    & 00 32 43.79 & -63 01 53.0 &  5.07 &  21.52$\pm$0.49 &    86.15$\pm$0.39 &   -49.85$\pm$0.46 &   5.0$\pm$1.2 &   16.3 &  -22.0 &  -37.6 &  -10.4$\pm$   0.5 &  -19.9$\pm$   0.7 &    1.0$\pm$   1.0 \\ 
HIP2729    & 00 34 51.09 & -61 54 57.7 &  9.56 &  21.78$\pm$1.01 &    87.33$\pm$0.89 &   -53.14$\pm$0.96 &  -1.0$\pm$3.0 &   15.6 &  -21.2 &  -37.7 &  -12.1$\pm$   1.2 &  -17.7$\pm$   1.6 &    6.2$\pm$   2.5 \\       
HIP9141    & 01 57 48.91 & -21 54 04.9 &  8.07 &  23.61$\pm$1.03 &   103.86$\pm$0.98 &   -50.89$\pm$0.75 &   5.7$\pm$0.3 &  -11.1 &   -3.5 &  -40.7 &  -10.4$\pm$   0.4 &  -21.5$\pm$   0.9 &   -1.3$\pm$   0.3 \\ 
HIP12394   & 02 39 35.22 & -68 16 01.0 &  4.12 &  21.27$\pm$0.50 &    87.40$\pm$0.42 &     0.56$\pm$0.50 &   6.4$\pm$2.2 &   10.6 &  -31.0 &  -33.7 &  -11.0$\pm$   0.6 &  -17.0$\pm$   1.5 &    3.2$\pm$   1.6 \\ 
HIP16853   & 03 36 53.32 & -49 57 28.9 &  7.62 &  24.00$\pm$0.66 &    89.96$\pm$0.66 &     1.82$\pm$0.70 &  14.4$\pm$0.9 &   -4.4 &  -25.8 &  -32.5 &  -10.1$\pm$   0.3 &  -20.5$\pm$   0.6 &   -0.8$\pm$   0.8 \\ 
HIP21965   & 04 43 17.17 & -23 37 41.9 &  7.12 &  17.17$\pm$1.24 &    50.00$\pm$0.81 &   -13.28$\pm$0.93 &  19.3$\pm$2.9 &  -33.7 &  -31.2 &  -35.8 &  -11.5$\pm$   1.7 &  -20.9$\pm$   1.7 &   -2.3$\pm$   1.9 \\ 
HIP100751  & 20 25 38.85 & -56 44 05.6 &  1.94 &  17.80$\pm$0.70 &     7.71$\pm$0.58 &   -86.15$\pm$0.45 &   2.0$\pm$3.0 &   43.4 &  -15.0 &  -32.4 &   -6.4$\pm$   2.3 &  -22.2$\pm$   1.2 &   -1.7$\pm$   1.7 \\ 
HIP105388  & 21 20 49.93 & -53 02 02.3 &  8.65 &  21.81$\pm$1.17 &    30.00$\pm$1.16 &   -94.36$\pm$0.60 &  -1.6$\pm$0.2 &   32.0 &   -9.0 &  -31.6 &   -8.2$\pm$   0.4 &  -20.0$\pm$   1.1 &   -0.3$\pm$   0.2 \\ 
HIP107947  & 21 52 09.67 & -62 03 07.7 &  7.22 &  22.18$\pm$0.80 &    43.57$\pm$0.47 &   -91.84$\pm$0.49 &   1.4$\pm$0.6 &   28.1 &  -15.9 &  -31.5 &   -9.1$\pm$   0.5 &  -19.8$\pm$   0.7 &   -0.1$\pm$   0.4 \\ 
HIP108195  & 21 55 11.34 & -61 53 11.0 &  5.92 &  21.49$\pm$0.67 &    43.53$\pm$0.38 &   -90.56$\pm$0.42 &   1.0$\pm$3.0 &   28.8 &  -16.3 &  -32.7 &   -9.4$\pm$   1.9 &  -20.1$\pm$   1.2 &    0.3$\pm$   2.1 \\ 
HIP116748  & 23 39 39.35 & -69 11 44.1 &  8.17 &  21.64$\pm$1.32 &    79.04$\pm$1.15 &   -67.11$\pm$1.20 &   7.4$\pm$0.2 &   21.3 &  -23.5 &  -33.6 &   -9.5$\pm$   0.8 &  -21.9$\pm$   1.1 &   -0.9$\pm$   0.4 \\ 
\\
\hline
  \end{tabular}
\end{table}
\end{landscape}

\end{document}